\documentclass[12pt,tightenlines,eqsecnum,floats,aps,amsmath,amssymb,nofootinbib,superscriptaddress,showpacs]{revtex4-1}

\usepackage{dcolumn}
\usepackage{bm}
\usepackage[spanish,english]{babel}
\usepackage{amsfonts}
\usepackage{amssymb}
\usepackage{graphicx}
\usepackage[latin1]{inputenc}

\newcommand{\Ham}{\mathcal{H}}

\newcommand{\lp}{\left(}
\newcommand{\rp}{\right)}

\begin{document}

\title{ {\bf Hamiltonian Formulation of Palatini $f(R)$ theories \`{a} la Brans-Dicke}}
\author{Gonzalo J. Olmo}\email{gonzalo.olmo@uv.es}
\affiliation{ \footnotesize Departamento de F\'{i}sica Te\'{o}rica and IFIC, Centro Mixto Universidad de Valencia - CSIC.
  Facultad de F\'{i}sica, Universidad de Valencia, Burjassot-46100, Valencia, Spain  and \\
  Instituto de Estructura de la Materia, CSIC, Serrano 121, 28006 Madrid, Spain}

\author{Helios Sanchis-Alepuz}\email{hesana@uv.es}
\affiliation{ {\footnotesize Institut f\"{u}r Physik, Karl-Franzens-Universit\"{a}t Graz, Austria }}
        

\date{January $18^{th}$, 2011}

\begin{abstract}
We study the Hamiltonian formulation of $f(R)$ theories of gravity both in metric and in Palatini formalism using their classical equivalence with Brans-Dicke theories with a non-trivial potential. The Palatini case, which corresponds to the $\omega=-3/2$ Brans-Dicke theory, requires special attention because of new constraints associated with the scalar field, which is non-dynamical. We derive, compare, and discuss the constraints and evolution equations for the $w=-3/2$ and $w\neq -3/2$ cases. Based on the properties of the constraint and evolution equations, we find that, contrary to certain claims in the literature, the Cauchy problem for the $w=-3/2$ case is well-formulated and there is no reason to believe that it is not well-posed in general. 
\end{abstract}


\maketitle

\section{Introduction}

In recent years modified theories of gravity of the $f(R)$ type have been thoroughly studied mainly because of the expected role that they could play in the understanding of the cosmic speedup. For different choices of the Lagrangian $f(R)$, one can find cosmological self-accelerating solutions without the need for sources of dark energy \cite{ReviewI,FarSot08}. In this context, it was soon realized \cite{Vol03,PalCosmo} that once an $f(R)$ lagrangian is given, the field equations could be derived in two inequivalent ways depending on whether the connection is defined as the Levi-Civita connection of the metric (metric formalism) or whether it is seen as independent of the metric (Palatini formalism). The resulting equations are radically different from each other and only lead to the same dynamics in the particular case of $f(R)=R-2\Lambda$. \\

The main focus in the literature has been on $f(R)$ theories in metric formalism, the usual variational formalism, which leads to higher order equations for the metric. Higher curvature terms of the $f(R)$ type are well-known in the literature \cite{well-known} and naturally arise in perturbative quantization schemes \cite{Parker-Toms} and in low-energy effective actions of string theories \cite{strings}, which justifies the interest in this kind of theories. The Palatini formulation, however, was not known to make contact with {\it more fundamental} approaches and has only received a timid attention as compared with its nonidentical twin brother in metric formalism. This situation changed recently \cite{Olmo-Singh-09}, when it was shown that a certain Palatini Lagrangian could faithfully reproduce the effective dynamics of Loop Quantum Cosmology (LQC) \cite{LQC}, a Hamiltonian based approach to quantum cosmology inspired by the non-perturbative quantization techniques of Loop Quantum Gravity \cite{LQG}. Though the Hamiltonian description of metric $f(R)$ theories has been discussed several times and from different points of view in recent literature \cite{Deruelle:2009pu,Ezawa:2009rh,Dyer:2008hb}, a discussion of the Palatini case is still missing. The main purpose of this paper, therefore, will be the study of the Hamiltonian formulation of Palatini $f(R)$ theories. \\

The dynamics of $f(R)$ theories in metric formalism can be analyzed in several classically equivalent ways. Some authors treat them as higher derivative theories, whereas others interpret them as scalar-tensor theories. Using a covariant approach, one finds that the higher order derivatives that appear in the field equations always act on a particular combination of second-order derivatives of the metric, namely, the scalar curvature, $R$ (or a function of it). Taking the trace of the metric field equations one then finds an independent equation which only involves (a function of) $R$ and derivatives with respect to $R$ of functions of $R$. This fact allows one to reinterpret the theory as a scalar-tensor theory instead of as an intrinsic fourth order theory, which simplifies the mathematics and makes the physics of the problem a bit more transparent. The scalar-tensor representation of the theory can be put into correspondence with a kind of Brans-Dicke like theories with characteristic parameter $\omega=0$ and a non-trivial potential, which is intimately related to the form of the function $f(R)$: $V(\phi)=R df/dR-f(R)$ and $\phi\equiv df/dR$. This technical advantage facilitated the analysis of cosmological models and of solar system tests. The results, however, are not very optimistic for those models in which $f(R)$ is of the form $R$ plus infrared corrections \cite{Chiba,Olmo1,Others}. Nonetheless, there is still hope that models with {\it chameleon} properties might be viable \cite{chameleon}. From a Hamiltonian point of view, it has recently been shown that the higher-derivative and the scalar-tensor interpretations are related by a canonical transformation \cite{Deruelle:2009pu}. \\

In the Palatini formulation of $f(R)$ theories, metric and connection are seen as independent variables. Upon variation of the action, one finds that the connection satisfies a constraint equation, which can be easily solved in terms of the metric and the matter fields, whereas the field equations for the metric are second order, like in General Relativity (GR). The resulting theory has, surprisingly, the same number of degrees of freedom (dynamical fields) as GR. The role of the Lagrangian $f(R)$ is thus to change the way matter generates curvature by adding new matter contributions, which stem from the connection constraint equation, on the right hand side of Einstein equations. In the case of infrared corrected models proposed to explain the cosmic speedup, such terms lead to incompatibilities with observations of many kinds \cite{ReviewI,FarSot08,Lab,SolSyst,Olmo-2008b,Polytropes}. For ultraviolet corrected models, on the contrary, one finds the same agreement with observations as in GR for curvatures much smaller than the ultraviolet scale of the theory \cite{Olmo-2008b,BOSA09,MyTalks}. Furthermore, it has been shown that the equations of motion that follow from the effective Hamiltonian of Loop Quantum Cosmology for an isotropic cosmology can be derived from a covariant action of the $f(R)$ type in Palatini formalism \cite{Olmo-Singh-09}. This cosmology turns out to be non-singular and replaces the disturbing Big Bang singularity of GR by a cosmic bounce, which occurs at a density of order  the Planck density. This effect is seen as a success of the non-perturbative techniques of the loop quantization. The existence of a Lagrangian that reproduces the LQC dynamics is important not only because it establishes the covariance of that Hamiltonian theory but also because it shows that Palatini $f(R)$ theories and its generalizations  could be an appropriate framework to study aspects of the phenomenology of quantum gravity. More recently, it has been shown that bouncing cosmologies are quite generic in the Palatini $f(R)$ framework \cite{BOSA09}, that extended theories of the type  $f(R,R_{\mu\nu}R^{\mu\nu})$ also share that property \cite{OSAT09,Barragan-2010}, and that $f(^\beta{R})$ theories, where $^\beta{R}$ is the scalar curvature of the Barbero-Immirzi connection with parameter $\beta$, are classically equivalent to Palatini $f(R)$ theories \cite{Fatibene:2010yc}. \\  

Palatini $f(R)$ theories also admit a scalar-tensor representation {\it à la} Brans-Dicke \cite{Wang:2004pq,Olmo1}. In this case, however, the theory is characterized by the parameter $\omega=-3/2$ and a non-trivial potential, which has the same form as in the metric version of the theory . This particular value of $\omega$ is very special because it leads to a non-dynamical scalar field, which justifies why the theory has the same number of dynamical fields as GR. In this sense, it is well-known that the existence of non-dynamical fields in a theory results in constraints in its corresponding Hamiltonian formulation. However, though the Hamiltonian description of generic Brans-Dicke theories is very well known, the particular case of Palatini $f(R)$  ($\omega=-3/2$ and $V(\phi)\neq 0$) has not been studied yet. In addition, due to the interesting connections existing between Palatini $f(R)$ and LQC, which is a purely Hamiltonian based theory, we find necessary a careful analysis of the Hamiltonian description of Palatini $f(R)$ theories. A better understanding of the Hamiltonian theory will provide a new viewpoint for the comparison between the metric and Palatini versions of $f(R)$ theories and might also provide new insights on the possibility of finding effective descriptions for the dynamics of LQC and of more general Hamiltonian quantum gravity theories.\\ 
Since the field equations of Palatini and metric $f(R)$ theories are (classically) equivalent to the Brans-Dicke cases $\omega=-3/2$ and $\omega=0$, respectively, in this paper we consider the Hamiltonian formulation of Brans-Dicke theories with a non-trivial potential rather than that of the original $f(R)$ versions. This will allow for a more transparent comparison between metric and Palatini cases or, with more generality, between the case $\omega=-3/2$ and all the other Brans-Dicke cases $\omega\neq-3/2$. We will show that the canonical momentum associated to the Brans-Dicke scalar becomes degenerate in the case $w=-3/2$, whereas it remains an independent degree of freedom for any other value of the parameter $w$. The $w=-3/2$ case, therefore, must be handled with care as a constrained system \cite{PAMD}. We will see that its constraint and evolution equations share many similarities with the $w\neq-3/2$ case though it also has some important differences. We will use these similarities and differences to comment on the well-posedness of the Cauchy problem for $w=-3/2$, which has been criticized in the literature \cite{Far07,FarSot08} (see also \cite{Capo09,Far09}, and \cite{Capo-Vignolo} for a different opinion). \\

\indent Being mathematically rigorous, we would like to remark that for the original Palatini $f(R)$ one should consider the $3+1$ decomposition of the metric and of the connection independently, which complicates the analysis and the comparison with the metric version of the theory. However, all the applications of Palatini theories found in the recent literature (including its applications to LQC and nonsingular cosmologies) are restricted to the properties of their equations of motion and, therefore, can be translated to the Brans-Dicke case without further discussion. For these reasons, at this stage we find an unnecessary complication the explicit $3+1$ decomposition of the original Palatini $f(R)$ theory. Such decomposition, however, seems unavoidable in more general Palatini theories whose Lagrangian contains other scalars besides $R$ (such as $R_{\mu\nu}R^{\mu\nu}$, $R_{\mu\nu\alpha\beta}R^{\mu\nu\alpha\beta}$) for which an on-shell scalar-tensor representation is neither known nor likely to exist. \\

The content of the paper is organized as follows. In section $II$, we present the covariant formulation of $f(R)$ theories and describe their relation with Brans-Dicke theories. We then use the scalar-tensor representation to work out the $3+1$ decomposition of the action previous to the construction of the Hamiltonian. The next step is to compute the Hamiltonian for general Brans-Dicke theories with $w\neq -3/2$ and compare our result with the literature. The case $w=-3/2$ is worked out and compared with the general case $w\neq-3/2$ in section $V$. We derive, compare, and discuss the constraint and evolution equations in Sec. $VI$, where we also comment on the Cauchy problem. We conclude with a brief summary and conclusions.

\section{Covariant formulation}

Let us begin by defining the action of Palatini $f(R)$ theories
\begin{equation}\label{eq:Pal-Action}
S[{g},\Gamma ,\psi_m]=\frac{1}{2\kappa^2}\int d^4
x\sqrt{-{g}}f({R})+S_m[{g}_{\mu \nu},\psi_m]
\end{equation}
Here $f({R})$ is a function of ${R}\equiv{g}^{\mu \nu }R_{\mu \nu }(\Gamma )$, with $R_{\mu \nu }(\Gamma )$ given by
$R_{\mu\nu}(\Gamma )=-\partial_{\mu}
\Gamma^{\lambda}_{\lambda\nu}+\partial_{\lambda}
\Gamma^{\lambda}_{\mu\nu}+\Gamma^{\lambda}_{\mu\rho}\Gamma^{\rho}_{\nu\lambda}-\Gamma^{\lambda}_{\nu\rho}\Gamma^{\rho}_{\mu\lambda}$
where $\Gamma^\lambda _{\mu \nu }$ is the connection. The matter action $S_m$ depends on the matter fields $\psi_m$, the metric $g_{\mu\nu}$, which defines the line element $ds^2=g_{\mu\nu}dx^\mu dx^\nu$, and its first derivatives (Christoffel symbols). By construction, we assume that the matter action does not depend on the connection $\Gamma^\lambda _{\mu \nu }$. Varying (\ref{eq:Pal-Action}) with respect to the metric $g_{\mu\nu}$ we obtain
\begin{equation}\label{eq:met-var-P}
f_RR_{\mu\nu}(\Gamma)-\frac{1}{2}f(R)g_{\mu\nu}=\kappa ^2T_{\mu
\nu }
\end{equation}
where $f_R\equiv df/dR$. From this equation we see that the scalar $R$ is algebraically related to the energy-momentum tensor via the trace equation
\begin{equation}\label{eq:trace-P}
Rf_R-2f=\kappa ^2T,
\end{equation}
The solution to this algebraic equation will be denoted by $R=R(T)$. The variation of (\ref{eq:Pal-Action}) with respect to $\Gamma^\lambda _{\mu \nu }$ must vanish
independently of (\ref{eq:met-var-P}) and gives
\begin{equation}\label{eq:con-var}
\nabla_\rho  \left[\sqrt{-g}\left(\delta ^\rho _\lambda
f_Rg^{\mu \nu }-\frac{1}{2}\delta ^\mu _\lambda f_Rg^{\rho
\nu }-\frac{1}{2}\delta^\nu_\lambda f_Rg^{\mu
\rho}\right)\right]=0 \ ,
\end{equation}
where $f_R\equiv f_R(R[T])$ must be seen as a function of the matter terms by virtue of (\ref{eq:trace-P}). This equation is equivalent to $\nabla_\alpha  \left[\sqrt{-g}f_Rg^{\beta \gamma}\right]=0$ and leads to 
\begin{equation}\label{eq:Gamma-1}
\Gamma^\lambda_{\mu \nu }=\frac{t^{\lambda \rho
}}{2}\left(\partial_\mu t_{\rho \nu }+\partial_\nu
t_{\rho \mu }-\partial_\rho t_{\mu \nu }\right)
\end{equation}
where  $t_{\mu \nu }\equiv f_R g_{\mu \nu }$. Expressing the connection in terms of $f_R$ and $g_{\mu \nu }$, (\ref{eq:met-var-P}) becomes
\begin{eqnarray}\label{eq:Gmn}
R_{\mu \nu }(g)-\frac{1}{2}g_{\mu \nu }R(g)&=&\frac{\kappa
^2}{f_R}T_{\mu \nu }-\frac{R f_R-f}{2f_R}g_{\mu \nu
}-\nonumber\\&-&\frac{3}{2(f_R)^2}\left[\partial_\mu f_R\partial_\nu
f_R-\frac{1}{2}g_{\mu \nu }(\partial f_R)^2\right]+\nonumber \\
&+& \frac{1}{f_R}\left[\nabla_\mu \nabla_\nu f_R-g_{\mu \nu }\Box
f_R\right] \ .
\end{eqnarray}
One should note that all the $R, f,$ and $f_R$ terms on the right hand side of this system of equations are functions of the trace $T$ of the matter energy-momentum tensor. In vacuum, $T=0$, those equations boil down to 
\begin{equation}\label{eq:Gmn-vac}
R_{\mu \nu }(g)-\frac{1}{2}g_{\mu \nu }R(g)=-\Lambda_{eff}g_{\mu\nu}
\end{equation}
where $\Lambda_{eff}\equiv (R_0 f_{R_0}-f_0)/2f_{R_0}$ is evaluated at the constant value $R_0=R(T=0)$ and plays the role of an effective cosmological constant. 

\subsection{Equivalence with Brans-Dicke theories}

The field equations (\ref{eq:Gmn}) derived above can be rewritten in a more
compact and illuminating form by introducing the following definitions
\begin{eqnarray}\label{eq:phi=f_R}
\phi&\equiv& f_R\\
V(\phi)&\equiv& R(\phi)f_R-f(R(\phi)) \label{eq:V=rf_R-f}
\end{eqnarray}
where $\phi$ represents a scalar field and $V(\phi)$ its
potential. Note that in eq.(\ref{eq:V=rf_R-f}) we have assumed
invertible the relation between $R$ and $f_R(R)$ to obtain $R(f_R)\equiv R(\phi)$.
The equations of motion for the metric can then be expressed as
follows
\begin{eqnarray}\label{eq:Gab-ST}
R_{\mu \nu }(g)-\frac{1}{2}g_{\mu \nu }R(g)&=&
\frac{\kappa^2}{\phi}T_{\mu\nu}-\frac{1}{2\phi}g_{\mu\nu}V(\phi)+\nonumber\\
 &+&\frac{\omega}{\phi^2}\left[\partial_\mu\phi\partial_\nu\phi-\frac{1}{2}g_{\mu\nu}(\partial
 \phi)^2\right]+\nonumber\\
 &+&\frac{1}{\phi}\left[\nabla_\mu\nabla_\nu\phi-g_{\mu\nu}\Box \phi\right]
\end{eqnarray}
where $\omega$ takes the value $\omega=-3/2$. It is straightforward to
verify that $f(R)$ theories in metric formalism lead to a similar set of equations but with the replacement 
$\omega=0$ (see for instance \cite{Olmo1} for details). In that case, the trace equation is $3\Box f_R+Rf_R-2f=\kappa^2T$. \\

The equation of motion for the scalar field $\phi$ in both metric and Palatini formalisms is provided by the trace of the original field equations  and can be expressed as
\begin{equation}\label{eq:phi-ST}
(3+2\omega)\Box \phi +2V(\phi)-\phi \frac{dV}{d\phi}=\kappa^2T
\end{equation}
Remarkably, the field equations (\ref{eq:Gab-ST}) and (\ref{eq:phi-ST}) can be derived from the
following action
\begin{eqnarray} \label{eq:ST}
S[{g}_{\mu \nu},\phi,\psi_m]&=&\frac{1}{2\kappa ^2 }\int d^4
x\sqrt{-{g}}\left[\phi {R}({g})-\right.
\\&-&\left.\frac{\omega}{\phi}(\partial_\mu \phi\partial^\mu\phi)-V(\phi)
\right]+S_m[{g}_{\mu \nu},\psi_m]\nonumber
\end{eqnarray}
which represents a Brans-Dicke like scalar-tensor theory (in the usual metric variational formalism). Note that in the
original Brans-Dicke theory the potential term was absent, $V(\phi)=0$ \cite{BD61}. The absence of a potential term leads to inconsistencies when $\omega=-3/2$, since then (\ref{eq:phi-ST}) becomes ill defined unless all matter sources satisfy $T=0$. For non-trivial $V(\phi)$, that equation makes perfect sense, since then $\phi$ can be expressed as an algebraic function of the trace, $\phi=\phi(T)$, which becomes constant in vacuum (and wherever $T=$constant).

\section{Hamiltonian formulation}

Having obtained the (on-shell) equivalence between $f(R)$ theories and Brans-Dicke theories, we use the scalar-tensor representation to work out the $3+1$ decomposition of the action needed to construct the corresponding Hamiltonian. With regard to the field equations, this procedure leads to the same classical results as  with the original $f(R)$ theory. We choose to work with the equivalent scalar-tensor action because the $f(R)$ representation in the Palatini formalism would require considerable extra effort for the implementation of the $3+1$ slicing of the space-time and the analysis of the constraints \cite{Peldan94} due to the a priori independence of the connection. In the scalar-tensor form, the independent Palatini connection has already been factored out and appears in the form of a scalar field coupled to the curvature and characterized by $\omega=-3/2$ and a non-trivial potential $V(\phi)$. We will see that this specific value of the parameter $\omega$ is interesting in its own right, though we prefer to maintain its relation with $f(R)$ theories to emphasize the motivations that led us to study this particular case.\\
From now on we use lower-case latin letters to represent space-time indices. 

\subsection{3+1 decomposition}

The $3+1$ decomposition of the action (\ref{eq:ST}) proceeds in the usual way \cite{ADM}. We consider a foliation of the spacetime manifold $\mathcal{M}$ into hypersurfaces $\Sigma_T$ of simultaneity characterized by a function $T(x)=$constant, a normalized timelike co-vector $n_a\propto \partial_a T$ normal to this hypersurface, and a {\it shift} vector $N^a$ orthogonal to $n^a=g^{ab}n_b$. This allows us to construct a time flow vector $t^a=N n^a+N^a$, where $N$ is known as {\it lapse}, and decompose the metric in the form $g_{ab}=h_{ab}+s n_a n_b$. The parameter $s=\pm 1=n_a n^a$ tells us whether the signature is Euclidean, $s=1$, or Lorentzian $s=-1$.   
Elementary, though lengthy, manipulations allow us to express the Lagrangian density of (\ref{eq:ST}) as follows
\begin{eqnarray}\label{eq:Lagrangian}
\mathcal{L}&=&\frac{\sqrt{h}}{2\kappa^2}\left\{N\phi\lp R^{\lp3\rp}-s(K_{ab}K^{ab}-K^2)\rp+2h^{ab}D_aND_b\phi\right.\nonumber\\
 & &\left.-\frac{\omega}{N\phi}\lp N^2h^{ab}D_a\phi D_b\phi+s\lp\dot{\phi}-N^aD_a\phi\rp^2\rp\right.\nonumber\\
 & &\left.-2K\lp\dot{\phi}-N^aD_a\phi\rp-NV\lp\phi\rp\right\}
\end{eqnarray}
where $K_{ab}=h_a^c h_b^d\nabla_d n_c$ is the extrinsic curvature, $D_a\phi=h_a^b\nabla_b\phi$, $\dot\phi=t^a\partial_a\phi$, $^{(3)}R$ is the Ricci scalar of the $3-$metric $h_{ab}$, and we have used the following relations
\begin{eqnarray}
R^{(4)}&=&R^{(3)}-s\left[K_{ab}K^{ab}-({K_a}^a)^2+2\nabla_cJ^c\right] \\
J^c&=& n^c\nabla_a n^a-n^a\nabla_a n^c\\
NJ^c\nabla_c\phi&=&s h^{cd}D_c\phi D_d N+K \lp\dot{\phi}-N^a D_a\phi\rp\\
\sqrt{|g|}&=&N\sqrt{h}
\end{eqnarray}
The canonical variables of the theory are $(g_{ab},\phi)\equiv(N,N^a,h_{ab},\phi)$. The canonical momenta are defined by the following expressions 
\begin{eqnarray}
\Pi_N &=&\frac{\delta S}{\delta \dot{N}}=0 \ , \ \Pi_a=\frac{\delta S}{\delta \dot{N}^a}=0 \ , \\\
\Pi^{ab}&=&\frac{\delta S}{\delta \dot{h}_{ab}}=\\ &&-\frac{s\sqrt{h}}{2\kappa^2}\left[ \phi\lp K^{ab}-Kh^{ab}\rp-\frac{h^{ab}}{N}\lp\dot{\phi}-N^cD_c\phi\rp\right]\label{eq:momentumh}\nonumber \ , \ \\
\pi_\phi&=&\frac{\delta S}{\delta\dot{\phi}}=-\frac{s\sqrt{h}}{2\kappa^2}\lp 2K+\frac{2\omega}{N\phi}\lp\dot{\phi}-N^cD_c\phi\rp\rp\label{eq:momentumphi}
\end{eqnarray}

Like in GR, we immediately see that the momenta conjugated to $N$ and $N^a$ are constrained to vanish. On the other hand, from the combination of $\Pi_h\equiv h_{ab}\Pi^{ab}$ and $\pi_\phi$, we find that 
\begin{equation} \label{eq:extra-constraint}
\Pi_h-\phi\pi_\phi=\left(\frac{3+2w}{N}\right)\frac{s\sqrt{h}}{2\kappa^2}\left(\dot\phi-N^cD_c\phi\right)
\end{equation}
is also constrained to vanish when $\omega=-3/2$, which is the case associated with Palatini $f(R)$. At this point, it is useful to rewrite the Lagrangian  density $\mathcal{L}$ using the definition for $\Pi^{ab}$ to eliminate the explicit dependence on $K_{ab}$ from it. The result is 
\begin{eqnarray}\label{eq:Lagrangian2}
\mathcal{L}&=&\frac{\sqrt{h}}{2\kappa^2}\left[N\left\{\phi R^{\lp3\rp}-\frac{s}{\phi}\frac{(2\kappa^2)^2}{h}\lp \Pi^{ab}\Pi_{ab}-\frac{\Pi^2}{2}\rp\right\}\right.\nonumber \\
 &-&\frac{N\omega}{\phi}D_c\phi D^c\phi+2D_c\phi D^cN-NV(\phi) \\
&+&\left.(1+s)2K\lp\dot{\phi}-N^cD_c\phi\rp+(3-2 s w)\frac{\lp\dot{\phi}-N^cD_c\phi\rp^2}{2N\phi}\right] \nonumber
\end{eqnarray}
The first term in the last line of this expression vanishes in the Lorentzian case, $s=-1$, and the second term also vanishes when $s=-1$ if $\omega=-3/2$. When $(3+2 w)\neq0$, the combination $\lp\dot{\phi}-N^cD_c\phi\rp$ in the last line can be expressed in terms of the momenta using (\ref{eq:extra-constraint}). To proceed with the construction of the Hamiltonian one must have in mind the above constraints and apply Dirac's algorithm for constrained Hamiltonian systems. From now on we take $s=-1$.

\subsection{General case $\omega\neq -3/2$}
 
In the general case $\omega\neq -3/2$, we have the same (primary) constraints as in GR, namely, $C_N\equiv\Pi_N(t,x)=0$ and $C_a\equiv\Pi_a(t,x)=0$. The Hamiltonian is constructed by introducing Lagrange multiplier fields $\lambda_N(t,x)$, $\lambda_a(t,x)$ for the primary constraints and performing the Legendre transform as usual with respect to the remaining {\it velocities}. The result is 
\begin{equation}
H=\int d^3x \left[\lambda_N C_N+\lambda^a C_a+N^a\Ham_a+N\Ham_N\right] \label{eq:H}
\end{equation}
where 
\begin{eqnarray}
C_N&=& \Pi_N \ , \ C_a=\Pi_a \ , \\
\Ham_N&=& \left(\frac{\sqrt{h}}{2\kappa^2}\right)\left[-\phi R^{(3)}+\frac{(2\kappa^2)^2}{h\phi}\left(\Pi^{ab}\Pi_{ab}-\frac{\Pi_h^2}{2}\right)+\right.\\
& & \left.\frac{\omega}{\phi}D_c\phi D^c\phi+V(\phi)+\frac{(2\kappa^2)^2}{2h\phi(3+2\omega)}\left(\Pi_h-\phi\pi_\phi\right)^2\right] \ ,  \label{eq:H_N}\nonumber\\
\Ham_a&=&-2h_{ab}D_c\Pi^{bc}+\pi_\phi D_a\phi  \label{eq:H_Na}
\end{eqnarray}
These expressions reproduce previous results found in the literature for Brans-Dicke theories with $V(\phi)=0$ \cite{Garay-GBellido}. For the dynamics to be consistent, the constraints must be preserved under evolution, which requires that $\dot{C}_N\equiv\{H,C_N\}=0$ and $\dot{C}_a\equiv\{H,C_a\}=0$, where the poisson bracket at time $t$ is defined as 
\begin{equation}
\{A(x),B(x')\}=\int d^3\sigma\left[\frac{\delta A(x)}{\delta \Pi^i(\sigma)}\frac{\delta B(x')}{\delta Q_i(\sigma)}-\frac{\delta B(x')}{\delta \Pi^i(\sigma)}\frac{\delta A(x)}{\delta Q_i(\sigma)}\right] \ ,
\end{equation}
where $\Pi^i$ and $Q_i$ generically represent the canonical variables. In particular, this definition leads to $\{\Pi^{ab}(t,\vec{x}),h_{cd}(t,\vec{x}'\}=\delta^a_{(c}\delta^b_{d)}\delta^{(3)}(\vec{x}-\vec{x}')$. By direct evaluation, one finds that $\dot{C}_N=-\delta H/\delta N=-\Ham_N$ and $\dot{C}_a=-\delta H/\delta N^a=-\Ham_a$. We thus see that on consistency grounds we must impose the secondary constraints $\Ham_N=0$ and $\Ham_a=0$, which implies that the Hamiltonian $H$ is constrained to vanish, like in GR. If matter is present, one must add the corresponding pieces $\delta H_{matt}/\delta N$ and $\delta H_{matt}/\delta N^a$ to these constraints, which leads to
\begin{eqnarray}
\label{eq:constrN}
&-&\phi R^{\lp 3\rp}+\frac{1}{\phi}\lp\tilde{\Pi}^{ab}\tilde{\Pi}_{ab}-\frac{\tilde{\Pi}^2}{2}\rp+\frac{\omega}{\phi}D_c\phi D^c\phi\\
&+&2h^{cd}D_cD_d\phi+V(\phi)+\frac{\left(\tilde{\Pi}_h-\phi\tilde{\pi}_\phi\right)^2}{2\phi(3+2\omega)}+\frac{1}{\alpha}\frac{\delta\mathcal{H}_{matt}}{\delta N}=0\nonumber \\
&-&2D_d\tilde{\Pi}_a^d+\tilde{\pi}_\phi D_a\phi+\frac{1}{\alpha}\frac{\delta\mathcal{H}_{matt}}{\delta N^a}=0 \label{eq:constrNa}
\end{eqnarray}
where we have defined $\alpha\equiv h^{1/2}/(2\kappa^2)$ and used the tilde to denote the tensorial quantities $\tilde{\pi}_\phi=\pi_\phi/\alpha$ and $\tilde{\Pi}^{ab}=\Pi^{ab}/\alpha$.  The evolution equations can be written as follows (this requires some lengthy calculations which we omit here)
\begin{eqnarray}
\dot{\phi}&=& N^aD_a\phi-\frac{N}{3+2w}\left(\tilde{\Pi}_h-\phi\tilde{\pi}_\phi\right) \\
\dot{\tilde{\pi}}_\phi&=& N\left[R^{(3)}+\frac{\tilde{\Pi}^{ab}\tilde{\Pi}_{ab}}{\phi^2}+\frac{w}{\phi^2}D_c\phi D^c\phi-\frac{dV}{d\phi}\right]\nonumber \\
&+&2w D_c\left(\frac{N}{\phi} D^c\phi\right)-2\Delta N+N^cD_c\tilde{\pi}_\phi \nonumber \\ &+&\frac{N(\tilde{\Pi}_h-\phi\tilde{\pi}_\phi)(\tilde{\Pi}_h-2\phi\tilde{\pi}_\phi)}{2\phi^2(3+2w)} \\
\dot{h}_{ab}&=&2D_{(a}N_{b)}+\frac{2N}{\phi}\left(\tilde{\Pi}_{ab}-\frac{\tilde{\Pi}_h}{2}h_{ab}\right)+\frac{N(\tilde{\Pi}_h-\phi\tilde{\pi}_\phi)}{\phi(3+2w)}h_{ab}\\
\dot{\tilde{\Pi}}^{ab}&=&-N\left[\phi \  ^{(3)}G^{ab}-\frac{w}{\phi}\left(D^a\phi D^b\phi-\frac{1}{2}h^{ab}D_c\phi D^c\phi\right)\right.\nonumber \\
&+&\left.\frac{2}{\phi}\left(\tilde{\Pi}^{ac}\tilde{\Pi}^b_c-\frac{h^{ab}}{4}\tilde{\Pi}^{mn}\tilde{\Pi}_{mn}\right)-\frac{\tilde{\Pi}_h}{2\phi}\left(3\tilde{\Pi}^{ab}-\frac{\tilde{\Pi}_h}{2}h^{ab}\right)\right] \nonumber\\
&+&N^cD_c \tilde{\Pi}^{ab}-\tilde{\Pi}^{ca}D_c N^{b}-\tilde{\Pi}^{cb}D_c N^{a}\nonumber \\
&+& D^aD^b(N\phi)-h^{ab}\Delta(N\phi)-2D^a N D^b \phi+h^{ab}D_cND^c\phi \nonumber \\
&-&\frac{NV}{2}h^{ab}-\frac{1}{\alpha}\frac{\delta H_{matt}}{\delta h_{ab}}-\frac{5N(\tilde{\Pi}_h-\phi\tilde{\pi}_\phi)}{2\phi(3+2w)}\tilde{\Pi}^{ab} \ ,
\end{eqnarray}
where we have denoted $h^{ab}D_aD_b f=\Delta f$. The equations for $\dot{\tilde{\pi}}_\phi$ and $\dot{\tilde{\Pi}}^{ab}$ have been obtained from those corresponding to $\dot{\pi}_\phi$ and $\dot{\Pi}^{ab}$ by using the generic relation  $\dot{\tilde{\pi}}=\dot{\pi}/\alpha-\tilde{\pi}q^{ab}\dot{q}_{ab}/2$, where $\dot{\pi}=\{H,\pi\}$. Using the tensorial quantities $\tilde{\pi}$ instead of the densities $\pi$, we remove the term $\alpha$ from the constraint and evolution equations everywhere except in the contributions coming from the matter Hamiltonian.

\subsection{Palatini $f(R)$ case: $\omega= -3/2$}

As we saw above, the case $\omega=-3/2$ has an additional constraint not present in the general case $\omega\neq -3/2$. Equation (\ref{eq:extra-constraint}) puts forward the fact that the momentum $\pi_\phi$ of the Brans-Dicke scalar $\phi$ can be expressed as a linear combination of the momenta associated to the $3-$metric $h_{ab}$. Taking the Lagrangian (\ref{eq:Lagrangian2}) particularized to $\omega=-3/2$ and following Dirac's algorithm, the Hamiltonian of this case becomes
\begin{equation}
\bar{H}=\int d^3x \left[\lambda_N C_N+\lambda^a C_a+\lambda_\phi C_\phi+N^a\Ham_a+N\bar{\Ham}_N\right] \label{eq:H-Pal}
\end{equation}
where 
\begin{eqnarray}
C_\phi&=&\Pi_h-\phi\pi_\phi \ , \\
\bar{\Ham}_N&=& \left(\frac{\sqrt{h}}{2\kappa^2}\right)\left[-\phi R^{(3)}+\frac{(2\kappa^2)^2}{h\phi}\left(\Pi^{ab}\Pi_{ab}-\frac{\Pi_h^2}{2}\right)+\right.\\
& & \left.\frac{\omega}{\phi}D_c\phi D^c\phi+V(\phi)\right] \ , \label{eq:H_N_Pal}
\end{eqnarray}
and the other constraints are the same as in the general case $w\neq -3/2$. The evolution equations are now given by
\begin{eqnarray}\label{eq:dphi-Pal}
\dot{\phi}&=& N^aD_a\phi-\lambda_\phi \phi \\
\dot{\tilde{\pi}}_\phi&=& N\left[R^{(3)}+\frac{\tilde{\Pi}^{ab}\tilde{\Pi}_{ab}}{\phi^2}+\frac{w}{\phi^2}D_c\phi D^c\phi-\frac{dV}{d\phi}\right] \label{eq:dpi_phi-Pal}\\
&-& 2\Delta N+2wD_c\left(\frac{N}{\phi} D^c\phi\right)+N^aD_a\tilde{\pi}_\phi -\frac{\lambda_\phi\tilde{\pi}_\phi}{2} \nonumber\\
\dot{h}_{ab}&=&2D_{(a}N_{b)}+\frac{2N}{\phi}\left(\tilde{\Pi}_{ab}-\frac{h_{ab}}{2}\tilde{\Pi}_h\right)+\lambda_\phi h_{ab}\label{eq:dhab-Pal}\\
\dot{\tilde{\Pi}}^{ab}&=&-N\left[\phi \  ^{(3)}G^{ab}-\frac{w}{\phi}\left(D^a\phi D^b\phi-\frac{1}{2}h^{ab}D_c\phi D^c\phi\right)\right.\nonumber \\
&+&\left.\frac{2}{\phi}\left(\tilde{\Pi}^{ac}\tilde{\Pi}^b_c-\frac{h^{ab}}{4}\tilde{\Pi}^{mn}\tilde{\Pi}_{mn}\right)-\frac{\tilde{\Pi}_h}{2\phi}\left(3\tilde{\Pi}^{ab}-\frac{\tilde{\Pi}_h}{2}h^{ab}\right)\right] \nonumber\\
&+&N^cD_c \tilde{\Pi}^{ab}-\tilde{\Pi}^{ca}D_c N^{b}-\tilde{\Pi}^{cb}D_c N^{a}\nonumber \\
&+& D^aD^b(N\phi)-h^{ab}\Delta(N\phi)-2D^a N D^b \phi+h^{ab}D_cND^c\phi \nonumber \\
&-&\frac{NV}{2}h^{ab}-\frac{1}{\alpha}\frac{\delta H_{matt}}{\delta h_{ab}}-\frac{5}{2}\lambda_\phi \tilde{\Pi}^{ab} \ \label{eq:dPiab-Pal}.
\end{eqnarray}    
Consistency of the evolution implies the following secondary constraints for $\dot{C}_N=0$ and $\dot{C}_a=0$
\begin{eqnarray}
\label{eq:constrN3/2}
&-&\phi R^{\lp 3\rp}+\frac{1}{\phi}\lp\tilde{\Pi}^{ab}\tilde{\Pi}_{ab}-\frac{\tilde{\Pi}^2}{2}\rp+\frac{\omega}{\phi}D_c\phi D^c\phi \nonumber\\
&+&2h^{cd}D_cD_d\phi+V(\phi)+\frac{1}{\alpha}\frac{\delta\mathcal{H}_{matt}}{\delta N}=0 \\
&-&2D_c\tilde{\Pi}_a^c+\tilde{\pi}_\phi D_a\phi+\frac{1}{\alpha}\frac{\delta\mathcal{H}_{matt}}{\delta N^a}=0 \label{eq:constrNa3/2} \ .
\end{eqnarray}
Using the evolution equations and the constraint (\ref{eq:constrN3/2}), one can verify that the evolution of $C_\phi$ leads to
\begin{eqnarray}
\dot{C}_\phi&=&\{\bar{H},C_\phi\}=\\
&-&2\alpha N V(\phi)+\alpha N\phi \frac{dV}{d\phi}-\frac{N}{2}\frac{\delta\mathcal{H}_{matt}}{\delta N}-h_{ab}\frac{\delta\mathcal{H}_{matt}}{\delta h_{ab}}\label{eq:Trace-constraint} \nonumber
\end{eqnarray}
Since $\dot{C}_\phi$ must vanish, we must impose the secondary constraint
\begin{equation}
\phi \frac{dV}{d\phi}-2V(\phi)-\frac{1}{2\alpha}\frac{\delta\mathcal{H}_{matt}}{\delta N}-\frac{1}{N\alpha}h_{ab}\frac{\delta\mathcal{H}_{matt}}{\delta h_{ab}}=0 \label{eq:const3/2}
\end{equation}
Using the definitions $T_{ab}=-\frac{2}{\sqrt{-g}}\frac{\delta\mathcal{L}_{matt}}{\delta g^{ab}}$ and $g^{ab}=h^{ab}-\frac{1}{N^2}\lp t^a-N^a\rp\lp t^b-N^b\rp$ to derive the relations
\begin{eqnarray}
\frac{\delta\mathcal{L}}{\delta N}&=&\frac{\delta\mathcal{L}}{\delta g^{cd}}\frac{\delta g^{cd}}{\delta N}=\frac{-2}{N}\lp g^{cd}-h^{cd}\rp\frac{\delta\mathcal{L}}{\delta g^{cd}}\\
 h^{ab}\frac{\delta\mathcal{L}}{\delta h^{ab}}&=&h^{ab}\frac{\delta\mathcal{L}}{\delta g^{ab}}\\
 T&=&-\frac{2}{\sqrt{-g}}\lp\frac{-N}{2}\frac{\delta\mathcal{L}_{matt}}{\delta N}+h^{ab}\frac{\delta\mathcal{L}_{matt}}{\delta h^{ab}}\rp \ ,
\end{eqnarray}
and the fact that $\frac{\delta\mathcal{L}_{matt}}{\delta N}=-\frac{\delta\mathcal{H}_{matt}}{\delta N}$ and $\frac{\delta\mathcal{L}_{matt}}{\delta h^{ab}}=-\frac{\delta\mathcal{H}_{matt}}{\delta h^{ab}}$, one can verify that (\ref{eq:const3/2}) yields
\begin{equation}
\phi \frac{dV}{d\phi}-2V(\phi)=\kappa^2T \ . \label{eq:T-Ham}
\end{equation}
This equation reproduces the relation (\ref{eq:phi-ST}) when $w=-3/2$ and establishes an algebraic relation between the trace of the energy-momentum tensor of matter and the scalar field $\phi=\phi(T)$. \\

\section{On the evolution and constraint equations}

Besides the number of constraints, the only difference between the Hamiltonians (\ref{eq:H}) and (\ref{eq:H-Pal}) is the last term appearing in the definition of $H_N$ in (\ref{eq:H_N}), which is proportional to $(\Pi_h-\phi\pi_\phi)^2$. That term stems from the last factor of (\ref{eq:Lagrangian2}), which obviously vanishes when $\omega=-3/2$ and, therefore, cannot appear in (\ref{eq:H_N_Pal}). For $w=-3/2$ one must introduce the constraint $C_\phi$ to account for the degeneracy imposed by the vanishing of $(\Pi_h-\phi\pi_\phi)$. This small difference in the Hamiltonians means that the constraint and evolution equations will share many terms in common. We have arranged the formulas in the text in a way that allows for a clear comparison between the $w=-3/2$ and the general $w\neq-3/2$ case. We thus see that the constraints $C_N=0$, $C_a=0$, and $\Ham_a=0$ are the same in both cases, whereas $\Ham_N=0$ and $\bar{\Ham}_N=0$ differ only slightly. Eq.(\ref{eq:constrN}) contains the term $\frac{\left(\tilde{\Pi}_h-\phi\tilde{\pi}_\phi\right)^2}{2\phi(3+2\omega)}$, which is absent in (\ref{eq:constrN3/2}). As a result, the relation between the evolution equations can be summarized as follows
\begin{eqnarray}
\dot{\phi}_{(w)}&=&\dot{\phi}_{(-\frac{3}{2})}+\lambda_\phi \phi-\frac{N}{3+2w}\left(\tilde{\Pi}_h-\phi\tilde{\pi}_\phi\right) \\
\dot{\tilde{\pi}}_\phi^{(w)}&=&\dot{\tilde{\pi}}_\phi^{(-\frac{3}{2})}+\frac{1}{2}\lambda_\phi \tilde{\pi}_\phi+\frac{N}{2\phi^2}\frac{\left(\tilde{\Pi}_h-\phi\tilde{\pi}_\phi\right)\left(\tilde{\Pi}_h-2\phi\tilde{\pi}_\phi\right)}{3+2w}\\
\dot{h}_{ab}^{(w)}&=&\dot{h}_{ab}^{(-\frac{3}{2})}-\lambda_\phi h_{ab}+\frac{N}{3+2w}\left(\tilde{\Pi}_h-\phi\tilde{\pi}_\phi\right)h_{ab}\\
\dot{\tilde{\Pi}}^{ab}_{(w)}&=&\dot{\tilde{\Pi}}^{ab}_{(-\frac{3}{2})}+\frac{5}{2}\lambda_\phi{\tilde{\Pi}}^{ab}-\frac{5N}{2\phi}\frac{\left(\tilde{\Pi}_h-\phi\tilde{\pi}_\phi\right)}{3+2w}{\tilde{\Pi}}^{ab}
\end{eqnarray}
These relations make it clear that there is no single (on-shell) choice of the function $\lambda_\phi$ that allows to recover the $w\neq-3/2$ equations from the $w=-3/2$ case, unless one makes the obvious substitution $\lambda_\phi\to \frac{(2\kappa^2)^2}{2h\phi(3+2\omega)}\left(\Pi_h-\phi\pi_\phi\right)$ directly in the Hamiltonian (\ref{eq:H-Pal}). \\ 

Now that we have compared the structure of the constraint and evolution equations, we will comment on their physical meaning and implications. 
In the case $w\neq -3/2$, the dynamical variables are $(\phi,\pi_\phi,h_{ab},\Pi^{ab})$ (plus the $(q_i,p^i)$ of the matter), while the lapse and shift,  $(N,N_a)$,  although necessary to determine the evolution, are undetermined by the equations. Note also that they do not enter into the constraint equations, for a change on the lapse-shift pair leaves the fields at the initial surface unchanged, which is a manifestation of the freedom to specify the coordinate system. In the $w=-3/2$ case, the dynamical variables are just $(h_{ab},\Pi^{ab})$ (plus the $(q_i,p^i)$ of the matter), because the evolution equations for $(\phi,\pi_\phi)$, as we saw above, can be combined to establish the secondary constraint (\ref{eq:T-Ham}). In this sense, it is worth noting that the constraint (\ref{eq:constrN}) involves up to second-order spatial derivatives of $h_{ab}$ and $\phi$ (see the terms $^{(3)}R$ and $D_aD_b\phi$), but only first order time derivatives of them (contained in the momenta $\Pi^{ab}$ and $\pi_\phi$). However, though the $w=-3/2$ constraint (\ref{eq:constrN3/2}) contains spatial derivatives of $\phi=\phi(T)$ up to second order, it does not contain any time derivative of $\phi(T)$ because the corresponding momentum $\pi_\phi$ is absent in that equation. Something analogous occurs in the vector constraint (\ref{eq:constrNa3/2}), where we can use the replacement $\pi_\phi=\Pi_h/\phi$ to show that no extra time derivatives of the matter appear in the constraints. This is a very important aspect, because it means that the highest order time derivative of the matter fields appearing in (\ref{eq:constrN3/2}) and (\ref{eq:constrNa3/2}) is the same as in GR and coincides with the highest order present in the energy-momentum tensor of the matter. The evolution equations also have this property. A glance at (\ref{eq:dphi-Pal}-\ref{eq:dPiab-Pal}) puts forward that the evolution equations for $\dot{\phi},\dot{\pi}_\phi,\dot{h}_{ab}$, and $\dot{\Pi}^{ab}$ do not contain the momentum $\pi_\phi$. Therefore, though one can find up to second-order spatial derivatives of $\phi(T)$, and hence of $T$, there is no trace of extra time derivatives acting on the matter fields. \\

It is worth noting that the second-order spatial derivatives of $\phi(T)$ require an extra degree of smoothness in the matter profiles, an aspect that is not necessary in GR. This extra degree of differentiability is a natural requirement if we attend to the $f(R)$ formulation of the $w=-3/2$ theory. Since the affine connection is compatible with a metric $t_{ab}$ which is conformally related with the space-time metric $g_{ab}$, the smoothness and differentiability of the conformal geometry is guaranteed if the conformal factor is differentiable up to second order (to yield a smooth field strength, Riemann tensor, of the affine connection). Since the conformal factor is the function $f_R(T)\equiv \phi(T)$, the differentiability condition on the geometry is transferred to the higher-order (spatial) differentiability of the matter fields living in that geometry. This should not come as a big surprise because modified theories of gravity usually lead to higher-order equations for the metric, which demand a higher degree of differentiability of the metric field. This, in particular, occurs with $f(R)$ theories in metric formalism. In the Palatini case, we do not find higher-order derivatives of the metric because the modified dynamics is due to the new role played by the matter fields. However, the existence of a conformal geometry intimately related with the matter fields ends up imposing the extra (spatial) differentiability conditions on the matter fields. \\
Now, the existence of up to second-order spatial derivatives  but not of time derivatives of ($\phi,\pi_\phi$) when $w=-3/2$ is an unsolved question that requires further study\footnote{In fact, since the field $\phi$ is not dynamical, the time evolution of the pair ($\phi,\pi_\phi$) must be completely provided by the time evolution of $(q_i,p^i)$, as is manifest from the secondary constraint (\ref{eq:T-Ham}). One could thus expect that the possible higher-order time derivatives of the matter fields induced by time derivatives of $\phi$ or $\pi_\phi$ in the field equations could be rewritten in terms of first-order time derivatives using the evolution equations for the matter variables $(q_i,p^i)$. However, as we are remarking here, no such higher-order time derivatives arise and, therefore, we do not need to worry about this problem. }. In this sense, we believe that a clearer understanding of the algebra of constraints in the case $w=-3/2$ could shed some useful light into this problem \cite{OSA-2011}.\\

\section{On the Cauchy problem}

It is well-known that if in GR one specifies initial values for $N, N^a, h_{ab}$ and $\Pi^{ab}$ which are consistent with the constraint equations, the evolution equations uniquely determine $h_{ab}$ and $\Pi^{ab}$, while $N$ and $N^a$ remain undetermined, which expresses the existing gauge freedom of the theory. This guarantees that the intrinsic (coordinate-independent) geometry of space-time is determined uniquely by an initial choice of $h_{ab}$ and $\Pi^{ab}$ \cite{Witten1962,ADM}. The same is true for the scalar-tensor theories considered here, thus implying that the initial value problem is well-formulated for all $w$. For the $w=-3/2$ case, the only difference with respect to GR is that one must specify an initial value for $\lambda_\phi$ and take also into account its corresponding constraint equation to consistently establish the initial data. \\
Using different variables and representations of the evolution and constraint equations, one can also proof the well-posedness of the initial value problem of GR and of generic Brans-Dicke theories with $w\neq -3/2$ in both Einstein and Jordan frames \cite{Salgado}. One can also make special choices for the lapse-shift pair and manipulate the corresponding $3+1$ equations of GR to show that the conjugate variables $h_{ab}$ and $\Pi^{ab}$ do satisfy a hyperbolic evolution system \cite{Choquet83}. Although we will not try here to explicitly demonstrate the well-posedness of the initial value problem for $w=-3/2$, we will exploit the resemblance between our constraint and evolution equations (\ref{eq:constrN3/2}),(\ref{eq:constrNa3/2}),(\ref{eq:T-Ham}) and (\ref{eq:dhab-Pal})-(\ref{eq:dPiab-Pal}) with those of GR to argue that the Cauchy problem is likely to be well-posed also for the Brans-Dicke case $w=-3/2$. \\

Note first that in vacuum, $T_{\mu\nu}=0$ or $H_{matt}=0$, the constraint (\ref{eq:T-Ham}) implies that $\phi$ is a constant, $\phi_0$, which turns the constraints (\ref{eq:constrN3/2}) and (\ref{eq:constrNa3/2}) into
\begin{eqnarray}
\label{eq:constrNGRL}
&-&\phi_0 R^{\lp 3\rp}+\frac{1}{\phi_0}\lp\tilde{\Pi}^{ab}\tilde{\Pi}_{ab}-\frac{\tilde{\Pi}^2}{2}\rp+V(\phi_0)=0 \\
&-&2D_c\tilde{\Pi}_a^c=0 \label{eq:constrNaGRL} \ .
\end{eqnarray}
With a simple constant rescaling of the metric, these constraints are the same as those of GR with a cosmological constant. Setting for consistency the Lagrange multiplier $\lambda_\phi=0$, the evolution equations for $h_{ab}$ and $\tilde{\Pi}^{ab}$ also recover the same form as those of GR with a cosmological constant. We can thus conclude that the Cauchy problem in vacuum is well-posed. \\
When matter is present, one should add to our system of equations those corresponding to the matter fields, which we assume standard. The strategy now would be to interpret the $\phi$-dependent terms, which are functions of the trace $T$, as part of a new (or modified) matter Hamiltonian. This way, the constraint and evolution equations maintain a structure that closely resembles that of GR except by some non-constant factors $\phi(T)$ that multiply or divide objects like $^{(3)}R$ and $\tilde{\Pi}^{ab}\tilde{\Pi}_{ab}$. If the matter fields satisfy the spatial differentiability requirements imposed by the constraint equations, the absence of higher-order time derivatives of the matter fields in the constraint and evolution equations suggests that the time evolution will be as well-posed as in GR. This, in fact, has been explicitly shown for a perfect fluid using the Einstein frame representation of the evolution equations \cite{Capo-Vignolo}. Obviously, since in general the well-posedness of the GR equations depends on the particular matter sources considered, the modification of the source terms induced by the existence of $\phi(T)$-dependent terms requires a model by model analysis. Therefore, though one cannot conclude that the Cauchy problem is well-posed for an arbitrary $f(R)$ Palatini Lagrangian, we find no reasons to suspect that it is ill-posed in general.\\

Before closing this section, we would like to mention that the Cauchy problem has been used in recent literature to criticize the viability of {\it all} Palatini $f(R)$ theories. In fact, in \cite{Far07} it was claimed that the disappearance of the d'Alambertian $\Box \phi$ from (\ref{eq:phi-ST}) for the value $w=-3/2$ implies that the non-dynamical field $\phi$ can be arbitrarily assigned on a region or on the entire spacetime, provided its gradient satisfies a degenerate equation [Eq. $(4.5)$ in that paper], which reduces to a constraint. This fact, it was stated, would make impossible to eliminate the term $\Box \phi$ from the evolution equations unless $\Box\phi=0$. This was interpreted as a {\it no-go theorem} for Palatini $f(R)$ gravity, which would  have an ill-formulated Cauchy problem even in vacuum. This interpretation is conceptually wrong because the scalar field in the $w=-3/2$ is just a given algebraic function of the trace $T$ and, therefore, is clearly specified by the matter content. Moreover, one should note that  Eq.$(4.5)$ of \cite{Far07} is not correct. That equation should recover the well-known relation $2V-\phi V'=\kappa^2T$ that establishes  the algebraic relation between $\phi$ and $T$ [our secondary constraint (\ref{eq:T-Ham})]. Using Eqs. $(3.4)$ and $(3.5)$ of \cite{Far07}, it is easy to check that the associated Eq. $(3.10)$ does recover our equation (\ref{eq:T-Ham}) in the Brans-Dicke case $w=-3/2$ (even though this is not the result obtained in \cite{Far07}) . This indicates that the first claims against the well-posedness of the Cauchy problem for Palatini $f(R)$ theories stemmed from the analysis of erroneous equations.\\
The strong conclusions of \cite{Far07} were a bit relaxed in \cite{FarSot08} (see in this sense \cite{Capo09,Far09}), where it was admitted that the Cauchy problem should be well-posed in vacuum and with radiation fields (for which $T=0$ and $\phi=$constant). In fact, in \cite{FarSot08} it was correctly noticed that in the $w=-3/2$ case the field $\phi$ could be algebraically solved in terms of $T$ (though their Eq. $(219)$ is the same as Eq.$(4.5)$ of \cite{Far07}). It was then argued that the existence of terms of the form $\Box \phi(T)$, which imply contributions of the form $\Box T$, would cause problems for the Cauchy problem. Though such terms and the possible existence of higher-order derivatives of the matter fields are certainly a reason for concern, it was prematurely concluded that the Cauchy problem for Palatini $f(R)$ theories was likely to be neither well-formulated nor well-posed unless the trace $T$ were constant. These conclusions contrast with our findings, which show that the evolution equations do not introduce higher-order time derivatives of the matter fields, which guarantees that the initial value problem is as well formulated as in GR \cite{Witten1962,ADM}.  \\

\section{Summary and Conclusions}

In this work we have carried out the Hamiltonian formulation of Brans-Dicke theories with a non-trivial potential paying special attention to the case $w=-3/2$, whose field equations are equivalent to those of an $f(R)$ theory in the Palatini formalism. We have found that the scalar field in the case $w=-3/2$ presents a degenerate momentum, which is proportional to a linear combination of the momenta of the induced metric $h_{ab}$. This degeneracy requires, following Dirac's algorithm for constrained systems, the introduction of a new constraint in the Hamiltonian. Consistency of the evolution of that constraint leads to a secondary constraint which establishes an algebraic relation between the scalar field and the trace of the energy-momentum tensor of the matter. We have written the constraint and evolution equations in a way that allows for a quick comparison between the general case $w\neq -3/2$ and $w=-3/2$. \\
We pointed out that the resulting constraint and evolution equations of the case $w=-3/2$ do not contain any higher-order time derivative of the matter fields, and only spatial derivatives of the scalar $\phi(T)$ appear up to second-order. This implies that the spatial profiles of the matter sources must satisfy stronger differentiability requirements than in GR. We have interpreted this property as a natural requirement due to the existence of a conformal geometry directly related with the matter fields.  By comparing the constraint and evolution equations of our theory with those of GR we have shown that the initial value Cauchy problem is well-formulated in general (because the intrinsic geometry of space-time is determined uniquely by an initial choice of $h_{ab}$, $\Pi^{ab}$, plus the corresponding position and momenta of the matter on the initial Cauchy surface) and have found reasons to believe that it is likely to be also well-posed, which contrasts with other opinions found in the literature. 
This comparison is possible because the equations do not contain time derivatives of either the scalar $\phi(T)$ nor of its momentum conjugate $\pi_\phi$. The reason for the absence of time derivatives of these objects is an unsolved question that may be related with particular properties of the enlarged algebra of constraints of the case $w=-3/2$, which will be studied elsewhere \cite{OSA-2011}.\\

{\bf Acknowledgements.} G.J.O. has been supported by the Spanish grants FIS2008-06078-C03-02, FIS2008-06078-C03-03, and the Consolider Programme CPAN (CSD2007-00042). The work of H.S.-A. was supported by by the Austrian Science Fund FWF
under Project No. P20592-N16. H.S.-A. is grateful to R. Alkofer for his support. G.J.O. thanks S. Capozziello for useful clarifications.


\begin{thebibliography}{99}




\bibitem{ReviewI}
A.~De Felice, S.~Tsujikawa,  Living Rev.\ Rel.\  {\bf 13}, 3 (2010);
S. Capozziello and M. Francaviglia,  Gen. Rel. Grav. 40, 357 (2008).

\bibitem{FarSot08}
V.Faraoni and T.P. Sotiriou, Rev. Mod. Phys. 82, 451 (2010).


\bibitem{Vol03}
D.N. Vollick, {\it Phys. Rev.}{\bf D} 68, 063510 (2003), astro-ph/0306630.

\bibitem{PalCosmo}
X.H. Meng, P. Wang, Class. Quant. Grav. 20, 4949 (2003); 
Class. Quant. Grav. 21, 951 (2004); Phys.Lett. B 584, 1 (2004);
S. Capozziello, V. F. Cardone, M. Francaviglia, Gen.Rel.Grav.38,711 (2006);
A. Borowiec, W. Godlowski, M. Szydlowski, Phys.Rev.D 74, 043502 (2006);
T. Koivisto, Phys. Rev. D 73, 083517 (2006);
M. Amarzguioui et al.,  A \& A 454, 707 (2006);
B.~Li, K.~-C.~Chan, M.~-C.~Chu, Phys.\ Rev.\  {\bf D76}, 024002 (2007);
F.~C.~Carvalho, E.~M.~Santos, J.~S.~Alcaniz and J.~Santos, JCAP {\bf 0809}, 008 (2008);
 T. Koivisto and H. Kurki-Suonio, Class. Quant. Grav. 23, 2355 (2006);
 S.Fay, R. Tavakol, and S. Tsujikawa, Phys. Rev. D 75, 063509 (2007);
S. Tsujikawa, K. Uddin, and R. Tavakol, Phys. Rev. {D 77}, 043007 (2008);


\bibitem{well-known}
J.D. Barrow and A.C. Ottewill, {\it J. Phys. A: Math. Gen.} {\bf 16} (1983) 2757-2776;
V. Müller, H.-J. Schmidt,{\it Gen. Rel. Grav.} {\bf 17} (1985) 769;
A. A. Starobinsky, H.-J. Schmidt, Class. Quant. Grav. 4 (1987) 695;
V. Müller, H.-J. Schmidt, A. A. Starobinsky, {\it Phys. Lett.} {\bf B} 202 (1988) 198.

\bibitem{Parker-Toms}
Parker L. and Toms D.J., {\it Quantum field theory in curved spacetime: quantized fields
and gravity}, (Cambridge University Press, Cambridge, England, 2009);
Birrel N.D. and Davies P.C.W., {\it Quantum fields in curved space}, (Cambridge University Press, Cambridge, England, 1982).

\bibitem{strings}
M. Green, J. Schwarz, and E. Wittem, ``Superstrings'', Cambridge University Press, Cambridge, England (1987). 

\bibitem{Olmo-Singh-09}
G.J. Olmo and P. Singh, JCAP 0901,  030 (2009).


\bibitem{LQC}
M. Bojowald, {\it Living Rev. Rel.} {\bf 8}, 11 (2005);
A. Ashtekar, {\it Nuovo Cim.} {\bf 122 B}, 135 (2007);
A. Ashtekar, T. Pawlowski and P. Singh, {\it Phys. Rev. Lett. } {\bf 96}, 141301 (2006); {\it Phys. Rev.} {\bf D} 73,  124038 (2006);
{\it Phys. Rev. } {\bf D} 74, 084003 (2006); 
L. Szulc, W. Kaminski, J. Lewandowski, {\it Class.Quant.Grav.} {\bf 24}, 2621 (2007); 
A. Ashtekar, T. Pawlowski, P.Singh, K. Vandersloot, {\it Phys.Rev.} {\bf D} 75, 024035 (2007); 
K. Vandersloot, {\it Phys.Rev.} {\bf D} 75, 023523 (2007);
E. Bentivegna and T. Pawlowski, arXiv:0803.4446 [gr-qc].


\bibitem{LQG}
A. Ashtekar and J. Lewandowski, {\it Class. Quant. Grav.} {\bf 21} (2004) R53; 
C. Rovelli, {\it Quantum Gravity}, (Cambridge U. Press, 2004); 
T. Thiemann, {\it Modern canonical quantum general relativity}, (Cambridge U. Press, 2007).

\bibitem{Deruelle:2009pu}
  N.~Deruelle, Y.~Sendouda and A.~Youssef,
  arXiv:0906.4983 [gr-qc].

\bibitem{Ezawa:2009rh}
  Y.~Ezawa, H.~Iwasaki, Y.~Ohkuwa, S.~Watanabe, N.~Yamada and T.~Yano,
  arXiv:0902.3317 [gr-qc].
\bibitem{Dyer:2008hb}
  E.~Dyer and K.~Hinterbichler,
  Phys.\ Rev.\  D {\bf 79} (2009) 024028
  [arXiv:0809.4033 [gr-qc]].


\bibitem{Chiba}
T. Chiba, Phys. Lett. B576, 5 (2003).

\bibitem{Olmo1}
G.J. Olmo, {\it Phys.Rev.} {\bf D} 75, 023511(2007); {\it Phys.Rev.Lett.} 95, 261102 (2005); {\it Phys.Rev.} {\bf D} 72, 083505 (2005).

\bibitem{Others}
A. L. Erickcek, T. L. Smith, and M. Kamionkowski, {\it Phys.Rev.} {\bf D} 74, 121501 (2006); 
Xing-Hua Jin, Dao-Jun Liu, and Xin-Zhou Li, arXiv:astro-ph/0610854v4.

\bibitem{chameleon}
J. A. R. Cembranos, {\it Phys.Rev.} {\bf D}73 (2006) 064029, gr-qc/0507039;
I. Navarro and K. Van Acoleyen, JCAP 0702 (2007) 022, gr-qc/0611127; 
W.Hu and I.Sawicki, arXiv:0705.1158;
T.Faulkner, M.Tegmark, E.F.Bunn, and Y.Mao, astro-ph/0612569.

\bibitem{Lab}
E. E. Flanagan, Phys. Rev. Lett.{\bf 92}, 071101 (2004);Class. Quant. Grav. 21, 3817 (2004);
G. J. Olmo, Phys. Rev. Lett. {\bf 98}, 061101 (2007); Phys. Rev. {\bf D 77}, 084021(2008);
B.~Li, D.~F.~Mota and D.~J.~Shaw, Phys. Rev. {\bf D 78}, 064018 (2008);  Class.\ Quant.\ Grav.\  {\bf 26}, 055018 (2009).

\bibitem{SolSyst}
G. J. Olmo, Phys. Rev. Lett. {\bf 95}, 261102(2005); Phys. Rev. {\bf D 72 }, 083505(2005);

\bibitem{Olmo-2008b}
G. J. Olmo, Phys. Rev. {\bf D 78}, 104026 (2008).

\bibitem{Polytropes}
K.Kainulainen, V. Reijonen, and D. Sunhede, {\it Phys.Rev.} D {\bf 76},043503(2007);
E. Barausse, T.P. Sotiriou, and J.C. Miller, {\it Class.Quant.Grav.} 25,062001(2008);
K.Kainulainen, J.Piilonen, V. Reijonen, and D. Sunhede, {\it Phys.Rev.} D {\bf 76},024020(2007);
E. Barausse, T.P. Sotiriou, and J.C. Miller, {\it Class.Quant.Grav.} 25,105008(2008);

\bibitem{BOSA09}
C.~Barragan, G.~J.~Olmo, H.~Sanchis-Alepuz,
  Phys.\ Rev.\  {\bf D80}, 024016 (2009).
  [arXiv:0907.0318 [gr-qc]];

\bibitem{MyTalks}
C.~Barragan, G.~J.~Olmo, H.~Sanchis-Alepuz, [arXiv:1002.3919 [gr-qc]];
 G.~J.~Olmo, H.~Sanchis-Alepuz, S.~Tripathi,  [arXiv:1002.3920 [gr-qc]];
 G.~J.~Olmo,  AIP Conf.\ Proc.\  {\bf 1241}, 1100-1107 (2010), [arXiv:0910.3734 [gr-qc]];


\bibitem{OSAT09}
  G.~J.~Olmo, H.~Sanchis-Alepuz and S.~Tripathi,
  Phys.\ Rev.\  D {\bf 80} (2009) 024013
  [arXiv:0907.2787 [gr-qc]].

\bibitem{Barragan-2010}
  C.~Barragan, G.~J.~Olmo,
  Phys.\ Rev.\  {\bf D82}, 084015 (2010).
  [arXiv:1005.4136 [gr-qc]].

\bibitem{Fatibene:2010yc}
  L.~Fatibene, M.~Ferraris, M.~Francaviglia,

  Class.\ Quant.\ Grav.\  {\bf 27}, 185016 (2010).
  [arXiv:1003.1619 [gr-qc]];  Class.\ Quant.\ Grav.\  {\bf 27}, 165021 (2010).
  [arXiv:1003.1617 [gr-qc]].

\bibitem{Wang:2004pq}
  P.~Wang, G.~M.~Kremer, D.~S.~M.~Alves {\it et al.},
  Gen.\ Rel.\ Grav.\  {\bf 38}, 517-521 (2006).
  [gr-qc/0408058].



\bibitem{PAMD}
Paul A.M. Dirac, {\it Lectures on Quantum Mechanics}, Belfer Graduate School of Science, Yeshiva University Press, New York, 1964.


\bibitem{Far07}
N. Lanahan-Tremblay and V. Faraoni, Class. Quant. Grav. 24, 5667 (2007).

\bibitem{Capo09}
S. Capozziello and S. Vignolo, Class. Quant. Grav. 26, 168001 (2009).

\bibitem{Far09}
V. Faraoni, Class. Quant. Grav. 26, 168002 (2009).

\bibitem{Capo-Vignolo}
  S.~Capozziello and S.~Vignolo, Class. Quant. Grav. 26, 175013 (2009) ;
  Int.\ J.\ Geom.\ Meth.\ Mod.\ Phys.\  {\bf 6}, 985 (2009)
  [arXiv:0901.3136 [gr-qc]].
  


\bibitem{BD61}
C. Brans and R. H. Dicke, {\it Phys. Rev.} 124, 925-935 (1961).



\bibitem{Peldan94}
P. Peldan, Class. Quant. Grav. {\bf 11}, 1087 (1994).



\bibitem{ADM}
C.W. Misner, S. Thorne, and J. A. Wheeler, {\it Gravitation}, (W.H. Freeman and Co., NY, 1973);

\bibitem{Wald1984}
Robert M. Wald, {\it General Relativity}, (The University of Chicago Press, Chicago, 1984).


\bibitem{Garay-GBellido}
L.J. Garay and J. Garcia-Bellido, Nucl.Phys. B 400, 416 (1993).

\bibitem{OSA-2011}
G.J. Olmo and H. Sanchis-Alepuz, work in progress. 

\bibitem{Witten1962}
L. Witten, ``Gravitation: an introduction to current research'', Ed. Wiley, New York (1962).


\bibitem{Salgado}
M. Salgado, Class. Quant. Grav. 23, 4719 (2006).

\bibitem{Choquet83}
Y. Choquet-Bruhat and T. Ruggeri, Commun. Math. Phys. 89, 269-275 (1983).


\end{thebibliography}
\end{document}